\documentclass[3p,times,twocolumn]{elsarticle}
\biboptions{numbers,compress}

\makeatletter
\def\ps@pprintTitle{%
     \let\@oddhead\@empty
     \let\@evenhead\@empty
     \let\@oddfoot\@empty
     \let\@evenfoot\@empty}
\makeatother

\newcommand{\ind}[2]{^{\mbox{\scriptsize $#1$}}_{\mbox{\scriptsize #2}}}
\newcommand{\Nc}{N_{\mbox{\scriptsize c}}}
\newcommand{\nf}{n_{\mbox{\scriptsize f}}}
\newcommand{\nfs}{n_{\mbox{\tiny f}}}
\newcommand{\eqnsgn}[1]{\!\!\!\!& #1 &\!\!\!\!}

\begin{document}

\begin{frontmatter}
\title{$R$--ratio of electron--positron annihilation into hadrons:
higher--order $\pi^2$--terms}
\author[Inst1]{A.V.~Nesterenko\fnref{fn1}}
\fntext[fn1]{Speaker, corresponding author}
\ead{nesterav@theor.jinr.ru}
\address[Inst1]{Bogoliubov Laboratory of Theoretical Physics,
Joint Institute for Nuclear Research,
Dubna, 141980, Russian Federation}
\author[Inst2]{S.A.~Popov}
\address[Inst2]{Moscow Institute of Physics and Technology,
Dolgoprudny, 141700, Russian Federation}
\begin{abstract}
High--energy behavior of $R$--ratio of electron--positron annihilation
into hadrons is studied. In~particular, it is argued that at any given
order of perturbation theory the re--expansion of the $R$--ratio in the
ultraviolet asymptotic can be reduced to the form of power series in the
naive continuation of the strong running coupling into the timelike
domain. The~convergence range of the re--expanded $R$--ratio and the
corresponding higher--order $\pi^2$--terms are discussed.
\end{abstract}
\begin{keyword}
electron--positron annihilation into hadrons \sep
timelike domain \sep
dispersion relations \sep
$\pi^2$--terms
\end{keyword}
\end{frontmatter}

The perturbative approach to Quantum Chromodynamics constitutes a basic
tool for the study of the strong interaction processes in the spacelike
(Euclidean) ultraviolet asymptotic\footnote{To~deal with the strong
interactions in the infrared domain one usually employs a variety of the
nonperturbative approaches to QCD, see, e.g., papers~\cite{Fried, HolQCD,
OPE, QCDLat, ILM, Frasca}.}. However, the theoretical description of
hadron dynamics in the timelike (Minkowskian) domain additionally requires
pertinent dispersion relations. The~latter not only allow one to handle
the strong interaction processes in the timelike domain in a
self--consistent way, but also provide intrinsically nonperturbative
constraints, which enable one to overcome some inherent difficulties of
the QCD~perturbation theory and extend its applicability range towards the
low energies, see papers~\cite{JPG32, PRD88JPG42} and references therein
for the details. It~is worth noting that the dispersion relations have
also proved their efficiency in such issues of theoretical particle
physics as the refinement of chiral perturbation theory~\cite{DRChPT},
accurate determination of parameters of resonances~\cite{DRRes},
assessment of the hadronic light--by--light scattering~\cite{Colangelo4},
as well as many others~\cite{APT0a, APT0b, APT1, APT2, APT3, APT4, APT5a,
APT5b, APT6, APT7a, APT7b, APT8, APT9, APT10}.

\begin{figure}[t]
\centerline{\includegraphics[width=77.5mm]{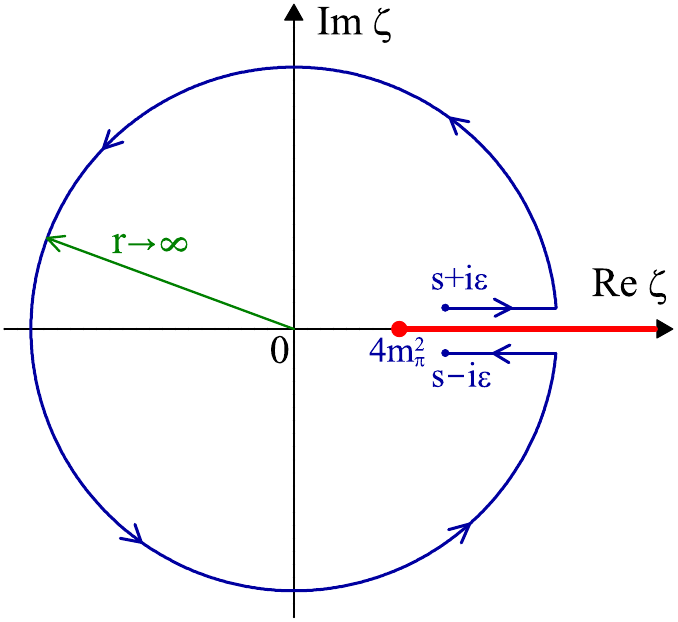}}
\caption{The~integration contour in Eq.~(\ref{AdlerInv}). The~physical cut
$\zeta \ge 4 m_{\pi}^2$ of the Adler function $D(-\zeta)$~(\ref{Adler_Def})
is shown along the positive semiaxis of real~$\zeta$.}
\label{Plot:Contour}
\end{figure}

Various strong interaction processes, including the hadronic contributions
to precise electroweak observables, involve the~hadronic vacuum
polarization function~$\Pi(q^2)$, which is defined as the scalar part of
the hadronic vacuum polarization tensor
\begin{eqnarray}
\label{P_Def}
\Pi_{\mu\nu}(q^2) \eqnsgn{=}
i\!\int\!d^4x\,e^{i q x} \bigl\langle 0 \bigl|
T\bigl\{J_{\mu}(x)\, J_{\nu}(0)\bigr\} \bigr| 0 \bigr\rangle =
\qquad
\nonumber \\
\eqnsgn{=} \frac{i}{12\pi^2}\, (q_{\mu}q_{\nu} - g_{\mu\nu}q^2)\, \Pi(q^2),
\end{eqnarray}
the corresponding $R(s)$~function
\begin{equation}
\label{R_Def}
R(s) = \frac{1}{\pi}\,{\rm Im}\lim_{\varepsilon \to 0_{+}}\!\Pi(s+i\varepsilon),
\qquad s = q^2 > 0,
\end{equation}
which is identified with the $R$--ratio of electron--positron annihilation
into hadrons, and the Adler function~\cite{Adler}
\begin{equation}
\label{Adler_Def}
D(Q^2) = - \frac{d\, \Pi(-Q^2)}{d \ln Q^2},
\qquad Q^2 = -q^2 > 0,
\end{equation}
where $Q^2 = -q^2 > 0$ and $s = q^2 > 0$ stand for the spacelike and
timelike kinematic variables, respectively. Though the function~$R(s)$ can
not be directly accessed within QCD~perturbation theory, it can be
expressed in terms of the Adler function~$D(Q^2)$ by making use of
Eqs.~(\ref{R_Def}) and~(\ref{Adler_Def}), specifically~\cite{Rad82, KP82}
\begin{equation}
\label{AdlerInv}
R(s) = \frac{1}{2 \pi i} \lim_{\varepsilon \to 0_{+}}
\int\limits_{s + i \varepsilon}^{s - i \varepsilon}
D(-\zeta)\,\frac{d \zeta}{\zeta}.
\end{equation}
The integration contour on the right--hand side of this equation lies in
the region of analyticity of the integrand, see Fig.~\ref{Plot:Contour}.

In~the ultraviolet asymptotic~$s\to\infty$ the effects due to the masses
of the involved particles\footnote{A~discussion of the impact of such
effects on the low--energy behavior of the functions on hand can be found
in papers~\cite{JPG32, PRD88JPG42, DPTPrelim, DPTAux} and references
therein.} can be safely neglected, so that in Eq.~(\ref{AdlerInv}) one can
employ the perturbative approximation of the Adler
function~\cite{AdlerPert4L}
\begin{equation}
\label{AdlerPert}
D^{(\ell)}_{{\rm pert}}(Q^2) = 1 + \sum_{j=1}^{\ell} d_{j}
\left[a\ind{(\ell)}{s}(Q^2)\right]^{j}
\end{equation}
(the~common prefactor $\Nc\sum_{f=1}^{\nfs} Q_{f}^{2}$ is omitted
throughout, where $\Nc=3$ stands for the number of colors, $Q_{f}$~denotes
the electric charge of $f$--th quark, and $\nf$~is the number of active
flavors). In~Eq.~(\ref{AdlerPert}) $d_j$~stand for the relevant
perturbative coefficients ($d_1 = 4/\beta_0$, $\beta_0 = 11 - 2\nf/3$),
whereas~$a\ind{(\ell)}{s}(Q^2)$ is the $\ell$--loop perturbative
QCD~couplant $a\ind{(\ell)}{s}(Q^2) = \alpha\ind{(\ell)}{s}(Q^2)\,
\beta_{0}/(4\pi)$, which satisfies the renormalization group
equation~($B_j=\beta_j/\beta^{j+1}_{0}$)
\begin{equation}
\label{RGPert}
\frac{d\,\ln\bigl[a\ind{(\ell)}{s}(\mu^2)\bigr]}{d\,\ln \mu^2} = -
\sum_{j=0}^{\ell-1} B_j
\left[a\ind{(\ell)}{s}(\mu^2)\right]^{j+1}.
\end{equation}
The~use of the perturbative approximation of the Adler
function~(\ref{AdlerPert}) in Eq.~(\ref{AdlerInv}) casts the latter to
(see Ref.~\cite{APT0b})
\begin{equation}
\label{RAPT}
R^{(\ell)}(s) = 1 +
\int\limits_{s}^{\infty}\!\rho^{(\ell)}_{{\rm pert}}(\sigma)\,
\frac{d \sigma}{\sigma},
\end{equation}
where
\begin{equation}
\label{RhoPert}
\rho^{(\ell)}_{{\rm pert}}(\sigma) \!=\!
\frac{1}{2 \pi i} \!\lim_{\varepsilon \to 0_{+}}
\Bigl[d^{(\ell)}_{{\rm pert}}(-\sigma - i \varepsilon) -
d^{(\ell)}_{{\rm pert}}(-\sigma + i \varepsilon)\!\Bigr]\;\,
\end{equation}
denotes the $\ell$--loop perturbative spectral function,
with~$d^{(\ell)}_{{\rm pert}}(Q^2)$ being the strong correction to the
Adler function~(\ref{AdlerPert}).

To~properly account for the effects due to continuation of the spacelike
perturbative results into the timelike domain one first has to
calculate\footnote{The~explicit expressions for the spectral
function~(\ref{RhoPert}) at various loop levels are given in
papers~\cite{Review, CPC} (see also Ref.~\cite{BCmath}).} the
spectral function~(\ref{RhoPert}) and then perform (explicitly or
numerically) the integration~(\ref{RAPT}). It~is worth noting that the
form of the resulting expression for the function~$R(s)$ drastically
differs from that of the perturbative power series~(\ref{AdlerPert}).
For~example, at the one--loop level Eq.~(\ref{RAPT}) reads (see
papers~\cite{Schr80, Rad82, AAP92, APT0b})
\begin{equation}
\label{RAPT1L}
R^{(1)}(s) = 1 +
d_{1}\! \left\{\frac{1}{2} -
\frac{1}{\pi}\arctan\!\left[\frac{\ln(s/\Lambda^2)}{\pi}\right]\right\}\!,
\end{equation}
where~$\Lambda$ is the QCD scale parameter and it is assumed
that~$\arctan(x)$ is a monotone nondecreasing function of its argument.
At~the same time, it appears that the re--expansion of Eq.~(\ref{RAPT}) in
the ultraviolet asymptotic~$s\to\infty$ leads to an approximate expression
for the function~$R(s)$, which resembles Eq.~(\ref{AdlerPert}).

In~particular, at high energies the $\ell$--loop strong correction to the
$R$--ratio~(\ref{RAPT}) can be represented as
\begin{equation}
r^{(\ell)}(s) =
\int\limits_{\ln w}^{\infty}\!\rho^{(\ell)}_{y}(y)\, d y, \qquad
y=\ln\left(\frac{\sigma}{\Lambda^2}\right)\!,
\end{equation}
where~$w = s/\Lambda^2$,
\begin{equation}
\label{RhoAppr}
\rho^{(\ell)}_{y}(y) = \frac{1}{2 \pi i}\,
\Bigl[d^{(\ell)}_{y}(y - i \pi) - d^{(\ell)}_{y}(y + i \pi)\Bigr],
\end{equation}
and~$d^{(\ell)}_{y}(y) = d^{(\ell)}_{{\rm
pert}}\bigl[\exp(y)\Lambda^2\bigr]$. Applying the Taylor expansion to
Eq.~(\ref{RhoAppr}) one arrives at
\begin{eqnarray}
\label{rAppr}
r^{(\ell)}_{{\rm pert}}(s) \eqnsgn{=} d^{(\ell)}_{{\rm pert}}(|s|) +
\nonumber \\
\eqnsgn{+}
\left.\sum_{n=1}^{\infty}
\frac{(-1)^{n} \pi^{2n}}{(2n+1)!}
\frac{d^{2n}}{d y^{2n}}\, d^{(\ell)}_{y}(y)\right|_{y=\ln w}.
\end{eqnarray}
The~right--hand side of this equation constitutes the sum of the naive
continuation~($Q^2 \to |s|$) of the strong correction to the Adler
function into the timelike domain (first line) and an infinite number of
the so--called \mbox{$\pi^2$--terms} (second line). Note that
Eq.~(\ref{rAppr}) is only valid for $\sqrt{s}>\exp(\pi/2)\Lambda$.

The~strong correction to the Adler function entering Eq.~(\ref{rAppr})
reads
\begin{equation}
d^{(\ell)}_{y}(y) = \sum_{j=1}^{\ell} d_{j}
\left[a^{(\ell)}_{y}(y)\right]^{j}\!,
\end{equation}
where~$a^{(\ell)}_{y}(y)=a\ind{(\ell)}{s}\bigl[\exp(y)\Lambda^2\bigr]$.
In~turn, Eq.~(\ref{RGPert}) implies
\begin{eqnarray}
\frac{d^{n}}{dy^{n}}\,\left[a^{(\ell)}_{y}(y)\right]^{j} \eqnsgn{=}
(-1)^{n}
\times \nonumber \\ \eqnsgn{} \hspace{-17.5mm}\times
\sum_{k_{1}=0}^{\ell-1}
\ldots
\sum_{k_{n}=0}^{\ell-1}
\,
\prod_{t=0}^{n-1}
\Bigl(j+t+k_{1}+\ldots+k_{t}\Bigr)
\times \nonumber \\ \eqnsgn{} \hspace{-17.5mm}\times
\prod_{p=1}^{n}B_{k_{p}}
\left[a^{(\ell)}_{y}(y)\right]^{j+n+
k_{1}+\ldots+k_{n}}\!,
\end{eqnarray}
that eventually leads to the following expression for the $\ell$--loop
perturbative approximation of \mbox{$R$--ratio} of electron--positron
annihilation into hadrons:
\begin{eqnarray}
\label{RPert1}
R^{(\ell)}_{{\rm pert}}(s) \eqnsgn{=} 1 +
\sum_{j=1}^{\ell} d_{j} \left[a^{(\ell)}_{{\rm s}}(|s|)\right]^{j} -
\nonumber \\ \eqnsgn{} \hspace{-17.5mm} -
\sum_{j=1}^{\ell} d_{j}
\sum_{n=1}^{\infty} \frac{(-1)^{n+1}\pi^{2n}}{(2n+1)!}
\sum_{k_{1}=0}^{\ell-1}
\ldots
\sum_{k_{2n}=0}^{\ell-1}\,
\prod_{p=1}^{2n}B_{k_{p}} \times
\nonumber \\ \eqnsgn{} \hspace{-17.5mm} \times\!
\prod_{t=0}^{2n-1}
\Bigl(j+t+k_{1}+\ldots+k_{t}\Bigr)\!
\left[a^{(\ell)}_{{\rm s}}(|s|)\right]^{j+2n+
k_{1}+\ldots+k_{2n}}\!\!. \quad
\end{eqnarray}
In~particular, this equation testifies that at any given loop level the
re--expansion of the $R$--ratio~(\ref{RAPT}) at high energies can be cast
to the form of power series in the naive continuation of the strong
running coupling into the timelike domain. Additionally,
Eq.~(\ref{RPert1}) explicitly demonstrates that the~$\pi^2$--terms appear
starting from the three--loop level only.

\begin{figure}[t]
\centerline{\includegraphics[width=77.5mm]{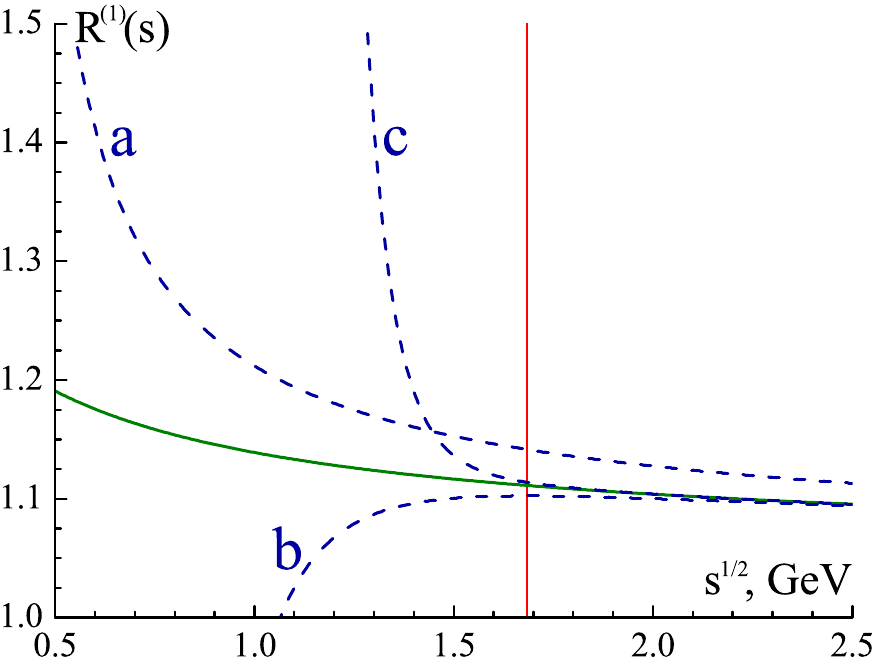}}
\caption{The one--loop function~$R^{(1)}(s)$~(\ref{RAPT1L}) (solid curve)
and its re--expansion~(\ref{R1LUV}) (dashed curves) truncated at the
orders~$\ln^{-1}z$ (curve~``a''), $\ln^{-7}z$ (curve~``b''),
and~$\ln^{-25}z$ (curve~``c''). Vertical line marks the lower boundary of
the convergence range of the re--expansion~(\ref{R1LUV}). The values of
parameters: $\nfs=3$ active flavors, $\Lambda=350\,$MeV.}
\label{Plot:R1LUV}
\end{figure}

Commonly, on the right--hand side of Eq.~(\ref{RPert1}) one discards
the~$\pi^2$--terms of the orders higher than the loop level on hand, that
results~in
\begin{equation}
\label{RPert2}
R^{(\ell)}_{{\rm pert}}(s) \simeq 1 + \!\sum_{j=1}^{\ell} r_{j}
\left[a\ind{(\ell)}{s}(|s|)\right]^{j}\!,
\quad r_{j} = d_{j} - \delta_{j}.
\end{equation}
It~is necessary to emphasize here that Eqs.~(\ref{RPert1})
and~(\ref{RPert2}) are only valid for~$\sqrt{s}>\exp(\pi/2)\Lambda \simeq
4.81\,\Lambda$, so that the lower boundary of the convergence range of the
re--expanded \mbox{$R$--ratio}~(\ref{RPert2}) may be as high as $(1.5
\ldots 2.0)\,$GeV for $\Lambda \simeq (300 \ldots 400)\,$MeV. This issue
is illustrated by Fig.~\ref{Plot:R1LUV}, which displays the one--loop
function~$R^{(1)}(s)$~(\ref{RAPT1L}) (solid curve), its re--expansion
\begin{eqnarray}
\label{R1LUV}
R^{(1)}(s) \eqnsgn{\simeq} 1 + d_{1} \biggl[
a\ind{(1)}{s}(|s|)
- \frac{\pi^2}{3}\frac{1}{\ln^3 w}
+ \frac{\pi^4}{5}\frac{1}{\ln^5 w}
- \nonumber \\
\eqnsgn{-} \frac{\pi^6}{7}\frac{1}{\ln^7 w}
+ \mathcal{O}\left(\frac{1}{\ln^9 w}\right)
\biggr],
\quad\; s \to \infty
\end{eqnarray}
truncated at various orders (dashed curves), and the lower boundary of the
convergence range of Eq.~(\ref{R1LUV}) (vertical~line).
In~Eq.~(\ref{R1LUV}) $a\ind{(1)}{s}(Q^2) = 1/\ln z$ stands for the
one--loop perturbative couplant, $z=Q^2/\Lambda^2$, and $w=s/\Lambda^2$.
In~particular, Fig.~\ref{Plot:R1LUV} shows Eq.~(\ref{R1LUV}) truncated at
the order~$\ln^{-1}z$, which corresponds to the naive continuation of the
one--loop perturbative approximation of the Adler
function~(\ref{AdlerPert}) into the timelike domain~$D^{(1)}_{{\rm
pert}}(|s|)$ (label~``a''), at the order~$\ln^{-7}z$ (label~``b''), and at
the order~$\ln^{-25}z$ (label~``c''). As~one can infer from this Figure,
the truncation of Eq.~(\ref{R1LUV}) at first order (that results in the
one--loop expression~(\ref{RPert2}) with~$\delta_1=0$) turns out to be a
rather rough approximation. Specifically, the relative difference between
the one--loop strong correction~(\ref{RPert2}), which basically discards
all the $\pi^2$--terms, and the one--loop strong
correction~(\ref{RAPT1L}), which, on the contrary, incorporates the
$\pi^2$--terms to all orders, appears to be about~$20\,\%$ at $\sqrt{s} =
2.5\,$GeV for $\nf=3$ active flavors and $\Lambda=350\,$MeV.

In~the perturbative approximation of the $R$--ratio~(\ref{RPert2}) the
coefficients~$\delta_{j}$, which embody the corresponding~$\pi^2$--terms,
have been calculated up to the sixth order in papers~\cite{Pi2Terms, KS95},
namely
\begin{equation}
\delta_{1} = 0,
\end{equation}
\begin{equation}
\delta_{2} = 0,
\end{equation}
\begin{equation}
\label{delta3}
\delta_{3} = \frac{\pi^2}{3} d_{1},
\end{equation}
\begin{equation}
\delta_{4} = \frac{\pi^2}{3} \left(\frac{5}{2} d_{1} B_{1} + 3 d_{2} \right)\!,
\end{equation}
\vskip-3mm
\begin{eqnarray}
\delta_{5} \eqnsgn{=}
\frac{\pi^2}{3} \left[ \frac{3}{2} d_{1} \left( B_{1}^{2} + 2 B_{2} \right)
+ 7 d_{2} B_{1} + 6 d_3 \right] -
\nonumber \\
\eqnsgn{-} \frac{\pi^4}{5} d_{1},
\end{eqnarray}
\begin{eqnarray}
\delta_{6} \eqnsgn{=}
\frac{\pi^2}{3} \biggl[ \frac{7}{2} d_{1} \left( B_{1} B_{2} + B_{3} \right)
+ 4 d_{2} \left( B_{1}^{2} + 2 B_{2} \right) +
\nonumber \\ \eqnsgn{+}
\frac{27}{2} d_3 B_{1} + 10 d_{4} \biggr]
- \frac{\pi^4}{5} \left( \frac{77}{12} d_{1} B_{1} + 5 d_{2} \right)\!.
\end{eqnarray}
The~higher--order coefficients~$\delta_j$~(\ref{RPert2}) read
\begin{eqnarray}
\delta_{7} \eqnsgn{=}
\frac{\pi^2}{3} \biggl[
4 d_{1} \left( B_{1} B_{3} + \frac{1}{2} B_{2}^{2} + B_{4} \right) +
\nonumber \\ \eqnsgn{+}
9 d_{2} \left( B_{1} B_{2} + B_{3} \right)
+ \frac{15}{2} d_{3} \left( B_{1}^{2} + 2 B_{2} \right) +
\nonumber \\ \eqnsgn{+}
22 d_{4} B_{1} + 15 d_{5} \biggr]
- \frac{\pi^4}{5} \biggl[
\frac{5}{6} d_{1} \left( 17 B_{1}^{2} + 12 B_{2} \right) +
\nonumber \\ \eqnsgn{+}
\frac{57}{2} d_{2} B_{1} + 15 d_{3} \biggr] + \frac{\pi^6}{7} d_1,
\end{eqnarray}
\begin{eqnarray}
\delta_{8} \eqnsgn{=}
\frac{\pi^2}{3} \biggl[
\frac{9}{2} d_{1} \left( B_{1} B_{4} + B_{2} B_{3} + B_{5} \right) +
\nonumber \\ \eqnsgn{+}
10 d_{2} \left(\! B_{1} B_{3} + \!\frac{B_{2}^{2}}{2}\! + B_{4}\! \right)
+ \!\frac{33}{2} d_{3} \left( B_{1} B_{2} + \! B_{3} \right) +
\nonumber \\ \eqnsgn{+}
12 d_{4} \left( B_{1}^{2} + 2 B_{2} \right) +
\frac{65}{2} d_{5} B_{1} + 21 d_{6} \biggr] -
\nonumber \\ \eqnsgn{-}
\frac{\pi^4}{5} \biggl[
\frac{15}{8} d_{1} \left( 7 B_{1}^{3} + 22 B_{1} B_{2} + 8 B_{3} \right) +
\nonumber \\ \eqnsgn{+}
\frac{5}{12} d_{2} \left( 139 B_{1}^{2} + 96 B_{2} \right)
\!+\! \frac{319}{4} d_{3} B_{1} + 35 d_{4}\! \biggr]\! +
\nonumber \\ \eqnsgn{+}
\frac{\pi^6}{7} \left( \frac{223}{20} d_{1} B_{1} + 7 d_{2} \right)\!.
\end{eqnarray}

\begin{figure}[t]
\centerline{\includegraphics[width=77.5mm]{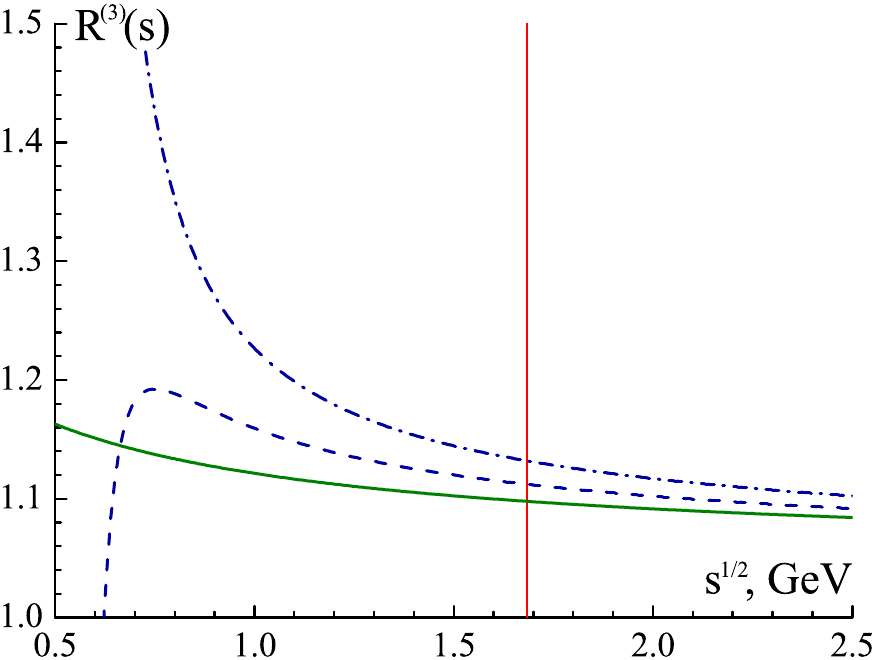}}
\caption{Three--loop function~$R^{(3)}(s)$ (Eq.~(\ref{RAPT}), solid
curve), its perturbative approximation~$R^{(3)}_{{\rm pert}}(s)$
(Eq.~(\ref{RPert2}), dashed curve), and the naive continuation of the
three--loop Adler function into the timelike domain~$D^{(3)}_{{\rm
pert}}(|s|)$ (Eq.~(\ref{AdlerPert}), dot--dashed curve). Vertical line
marks the lower boundary of the convergence range of the perturbative
approximation of $R$--ratio~(\ref{RPert2}). The~values of parameters:
$\nfs=3$ active flavors, $\Lambda=350\,$MeV.}
\label{Plot:RAPT3L}
\end{figure}

It~appears that the values of the coefficients~$\delta_j$ ($j \ge 3$)
substantially exceed the values of the corresponding perturbative
coefficients~$d_j$, that drastically affects the perturbative
approximation of $R$--ratio~(\ref{RPert2}). For~example, for $\nf=3$
active flavors $d_3=0.559$, whereas~$\delta_3=1.462$, so that the
third--order term in Eq.~(\ref{RPert2}) turns out to be amplified and even
sign--reversed~($r_3=-0.903$). This issue is illustrated by
Fig.~\ref{Plot:RAPT3L}, which displays the three--loop\footnote{The
scheme--dependent perturbative coefficients are assumed to be taken in the
$\overline{{\rm MS}}$--scheme throughout the paper.} function~$R^{(3)}(s)$
(Eq.~(\ref{RAPT}), solid curve), which includes the \mbox{$\pi^2$--terms}
to all orders, its perturbative approximation~$R^{(3)}_{{\rm pert}}(s)$
(Eq.~(\ref{RPert2}), dashed curve), which retains only the first
non--vanishing \mbox{$\pi^2$--term}~(\ref{delta3}), and the naive
continuation of the three--loop Adler function into the timelike
domain~$D^{(3)}_{{\rm pert}}(|s|)$ (Eq.~(\ref{AdlerPert}), dot--dashed
curve), which discards all the \mbox{$\pi^2$--terms}. The~lower boundary
of the convergence range of the perturbative approximation of
\mbox{$R$--ratio}~(\ref{RPert2}) at~\mbox{$\sqrt{s}=\exp(\pi/2)\Lambda$}
is shown in Fig.~\ref{Plot:RAPT3L} by vertical line. It~is worth noting
also that for~$\nf=3$ active flavors and~$\Lambda=350\,$MeV the relative
difference between the strong corrections $d^{(3)}_{{\rm
pert}}(|s|)$~(\ref{AdlerPert}) and~$r^{(3)}(s)$~(\ref{RAPT})
at~$\sqrt{s}=2.5\,$GeV exceeds~$20\,\%$, whereas the relative difference
between the strong corrections $r^{(3)}_{{\rm pert}}(s)$~(\ref{RPert2})
and~$r^{(3)}(s)$~(\ref{RAPT}) turns out to be about~$10\,\%$.

\smallskip

An evident advantage of the representation of the \mbox{$R$--ratio} of
electron--positron annihilation into hadrons in the form of
Eq.~(\ref{RAPT}) is that the latter properly accounts for the effects due
to continuation of the spacelike perturbative results into the timelike
domain and embodies (or~``resummates'') the $\pi^2$--terms to all orders.
The~integration in Eq.~(\ref{RAPT}) can easily be performed in the way
described in papers~\cite{Review, CPC}.

\end{document}